\pdfoutput=1 

\documentclass[%
 reprint,
 groupedaddress,
 showpacs, 
 noshowkeys,
 amsmath,amssymb,
 aps,
 prl,
]{revtex4-1}

\usepackage{graphicx}
\usepackage{dcolumn}
\usepackage{bm}
\usepackage{times}

\usepackage[cdot,mediumqspace,amssymb]{SIunits} 
\usepackage[autolanguage]{numprint} 

\usepackage{ifthen}
\usepackage{ifpdf}
\ifpdf
    \graphicspath{{Figures/PNG/}}
\else
    \graphicspath{{Figures/EPS/}}
\fi

\newcommand\eqnref[1]{Eq.~\eqref{#1}}
\newcommand\eqnsref[1]{Eqs.~\eqref{#1}}
\newcommand\figref[1]{Fig.~\ref{#1}}
\newcommand{\ket}[1]{\left| #1 \right>} 
\providecommand{\abs}[1]{\lvert#1\rvert}%

\def\A1{{\mathrm{A}_1}}
\def\Ex{{\mathrm{E}_x}}
\def\Ey{{\mathrm{E}_y}}
\def\dmupar{\Delta\mu_\parallel}
\def\muperp{\mu_\perp}
\def\meanf{\bar{\nu}}
\def\Vdc{V_\mathrm{dc}}
\def\rectvec{\boldsymbol{\xi}}
\def\dvec{\hat{\mathbf{v}}}

\begin{document}


\title{Electrical Tuning of Single Nitrogen-Vacancy Center \\ Optical Transitions Enhanced by Photoinduced Fields}

\author{L.\ C.\ Bassett}
 \thanks{These authors contributed equally to this work}
\author{F.\ J.\ Heremans}
 \thanks{These authors contributed equally to this work}
\author{C.\ G.\ Yale}
 \thanks{These authors contributed equally to this work}
\author{B.\ B.\ Buckley}
 \thanks{These authors contributed equally to this work}
\author{D.\ D.\ Awschalom}
 \email[Corresponding author.\\ Email address: ]{awsch@physics.ucsb.edu}
\affiliation{Center for Spintronics and Quantum Computation,\\ University of
California, Santa Barbara, California 93106, USA}

\date{\today}

\begin{abstract}
We demonstrate precise control over the zero-phonon optical transition
energies of individual nitrogen-vacancy (NV) centers in diamond by applying
multiaxis electric fields, via the dc Stark effect. The Stark shifts display
surprising asymmetries that we attribute to an enhancement and rectification
of the local electric field by photoionized charge traps in the diamond.
Using this effect, we tune the excited-state orbitals of strained NV centers
to degeneracy and vary the resulting degenerate optical transition frequency
by $>$\unit{10}{\giga\hertz}, a scale comparable to the inhomogeneous
frequency distribution. This technique will facilitate the integration of
NV-center spins within photonic networks.
\end{abstract}

\maketitle

Nitrogen-vacancy (NV) centers in diamond are promising solid-state qubits for
emerging quantum technologies, due to their long spin coherence times
\cite{Balasubramanian2009} and fast manipulation rates \cite{Fuchs2009},
together with a level structure that allows for straightforward optical
initialization and readout of the electronic spin state \cite{Manson2006}.
Furthermore, in high-quality single-crystal diamond at temperatures below
$\approx$\unit{25}{\kelvin} \cite{Fu2009}, sharp zero-phonon-line (ZPL)
optical transitions facilitate the coherent coupling between NV-center spins
and photons \cite{Buckley2010,Togan2010}. The integration of NV centers
within photonic structures \cite{Faraon2011,*Santori2010} to route single
photons and enhance the spin-photon interaction could therefore lead to
scalable applications for quantum information processing and secure
communication \cite{Barrett2005,*Childress2005}.

As solid-state ``trapped atoms,'' NV centers are sensitive to their local
environment. While this sensitivity has enabled nanoscale magnetic
\cite{Maze2008,*Balasubramanian2008,*Maertz2010} and electric
\cite{Dolde2011} metrology, it also exposes individual NV centers to sample
inhomogeneities, leading to a distribution of ZPL frequencies within a
diamond \cite{Batalov2009}. The ability to tune these frequencies is crucial
for photonic applications, for instance to utilize the selection rules at the
$C_{3v}$ symmetry point for spin-photon entanglement \cite{Togan2010} or to
coherently couple distant NV centers to indistinguishable photons. Through
the dc Stark effect, applied electric fields perturb both the ground-state
spin \cite{Dolde2011,VanOort1990} and excited-state orbitals
\cite{Tamarat2006a,*Tamarat2008}, providing the means to control the optical
transitions.

Here we use micron-scale devices to manipulate electric fields in three
dimensions, to compensate the intrinsic local strain and electrostatic fields
of individual NV centers and achieve full control of the orbital Hamiltonian.
Furthermore, by analyzing the Stark shifts as a function of applied voltages,
we infer a surprising amplification and rectification of the local electric
field, consistent with electrostatic contributions from photoionized charge
traps within the diamond host.  By harnessing this reproducible effect, we
can tune the NV-center Hamiltonian to arbitrary points across a range
comparable to the inhomogeneous ZPL distribution.

\begin{figure}[]
\includegraphics{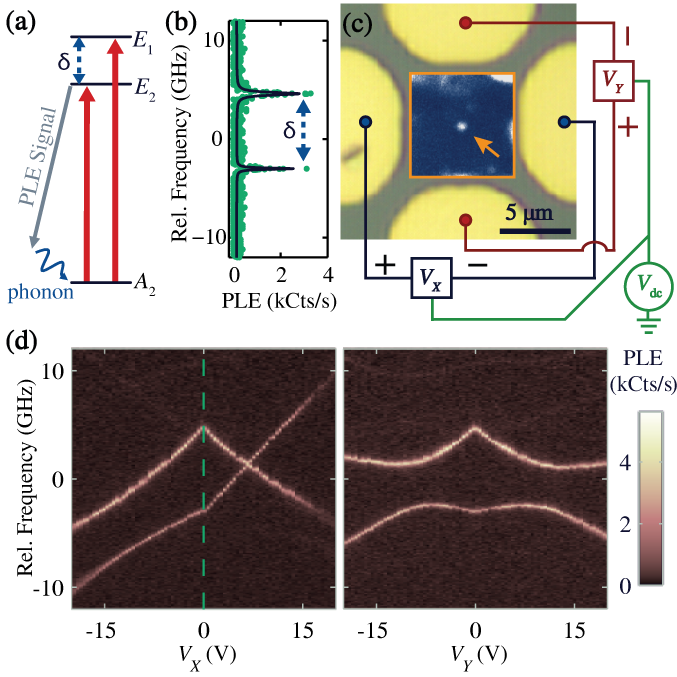}
  \caption[]{\label{fig:LateralExpt}
(a) Simplified energy-level diagram (not to scale) showing only the
$\ket{S_z}$ levels of the NV-center ground and excited states, with resonant
excitation (red arrows) and redshifted emission (gray arrow) marked. (b)
PLE spectrum (points) with no applied bias, marked by dashed line in (d),
with a two-Lorentzian fit (solid curve). (c) Micrograph and photoluminescence
image (center) of device A, with electrical connections marked. (d) PLE
spectra of the 6-\micro\metre-deep NV center marked by an arrow in
(c) as a function of lateral bias applied symmetrically to the $X$ (left
panel) or $Y$ (right panel) gate pairs, with $\Vdc=\unit{0}{\volt}$. In all
PLE spectra, the origin of the relative frequency axis is arbitrary.}
\end{figure}

The electronic structure of the negatively charged NV center is determined by
symmetry, through its point group $C_{3v}$
\cite{Batalov2009,Doherty2011,*Maze2011}. The spin-triplet ground (GS,
symmetry $^3\mathrm{A}_2$) and excited states (ES, symmetry $^3\mathrm{E}$)
are connected by ZPL transitions around \unit{637.2}{\nano\metre}
(\unit{1.946}{\electronvolt}).  Our experiments are performed at zero
magnetic field, where the spin-triplet basis states are
$\left\{\ket{S_x},\ket{S_y},\ket{S_z}\right\}$. A \unit{532}{\nano\metre}
(\unit{2.3}{\electronvolt}) ``repump'' beam pulsed at
$\approx$\unit{300}{\kilo\hertz} in a confocal geometry maintains a
spin-polarized population in $\ket{S_z}$. Between repump cycles, we count
photoluminescence excitation (PLE) photons emitted by the NV center into the
redshifted phonon sideband [see \figref{fig:LateralExpt}(a)] after absorption
from a narrow-line red laser tunable across the ZPL transitions. As we scan
the red laser frequency, we typically measure two peaks in the PLE spectrum
as shown in \figref{fig:LateralExpt}(b); these correspond to spin-conserving
transitions from the GS orbital singlet $\ket{\mathrm{A_2},S_z}$ to the two
ES orbital eigenstates $\left\{\ket{\mathrm{E_1},S_z},
\ket{\mathrm{E_2},S_z}\right\}$.
In a crystal environment with perfect $C_{3v}$ symmetry and zero electric and
magnetic fields, these ES orbital states would be degenerate, but the
symmetry is generally broken by local crystal strain and by nonuniform
electrostatic charge distributions that generate local electric fields.

The dc Stark  perturbation to the Hamiltonian, $\hat{H}_\mathrm{Stark} =
-\hat{\boldsymbol{\mu}}\cdot\mathbf{F}$, describes the interaction between
the local electric field $\mathbf{F}$ and the electric dipole operator
$\hat{\boldsymbol{\mu}}$.  For fixed stress, the strain perturbation can be
cast into the same form by isolating components which transform as the
irreducible representations of $C_{3v}$ \cite{Doherty2011,*Maze2011}.  The
combined perturbation has the form $V_\A1 \hat{O}_\A1 + V_\Ex \hat{O}_\Ex +
V_\Ey \hat{O}_\Ey$,  where $\hat{O}_{\Gamma_a}$ is an orbital operator
transforming as the basis state $\ket{\Gamma_a}$ and
\begin{equation}\label{eq:SymmetrizedFields}
  \left\{
  \begin{array}{l}
  V_\A1 = S_\A1 - \mu_\parallel F_z \\
  V_\Ex = S_\Ex - \mu_\perp F_x \\
  V_\Ey = S_\Ey - \mu_\perp F_y
  \end{array}\right.
\end{equation}
are the symmetrized field strengths, in terms of fixed strain components
$S_{\Gamma_a}$, projections of the local electric field $F_i$, and the
reduced matrix elements of the electric dipole operator
$\bigl\{\mu_\parallel,\mu_\perp\bigr\}$. We choose orbital basis states
$\left\{\ket{\mathrm{E}_x},\ket{\mathrm{E}_y}\right\}$ for the ES which
transform like vectors $\left\{x,y\right\}$ in the NV-center coordinate
system \footnote{The NV-center coordinate system is chosen such that
$\hat{\mathbf{z}}$ points along the N-V symmetry axis and $\hat{\mathbf{x}}$
lies in a reflection plane.}, and we ignore the small ($\approx$100~MHz)
spin-spin coupling between ES spin states $\ket{S_z}$ and
$\left\{\ket{S_x},\ket{S_y}\right\}$ \footnote{Note that while spin-spin
coupling leads to mixed ES spin eigenstates in some regimes, it does not
significantly affect the optical pumping mechanism which polarizes the spin
into $\ket{S_z}$ in our experiments.}. By defining
$\hat{H}\ket{\mathrm{A}_2,S_z}\equiv 0$, the Hamiltonian in the
$\left\{\ket{\mathrm{E}_x,S_z}, \ket{\mathrm{E}_y,S_z}\right\}$ basis can be
written as
\begin{equation}
  H = (\hbar\omega_0+\dmupar F_z)\mathbf{I}+
  \frac{1}{\sqrt{2}}\left(
\begin{array}{cc}
V_\Ex & -V_\Ey \\
-V_\Ey & -V_\Ex
\end{array}\right),
\end{equation}
where $\hbar\omega_0$ is the natural transition energy including fixed
perturbations of $\A1$ symmetry, and $\dmupar =
\bigl(\mu_\parallel^\mathrm{GS}-\mu_\parallel^\mathrm{ES}\bigr)$, defined
such that both $\dmupar$ and $\muperp$ are positive.  The transition energy
eigenvalues take the form $E_\pm = h\meanf \pm \frac{1}{2}h\delta$, where
\begin{subequations}  \label{eq:DCStarkComponents}
\begin{align}
  h\meanf & =  \hbar\omega_0+\dmupar F_z, \label{eq:meanfdefn}\\
  h\delta & = \sqrt{2}\left(V_\Ex^2+V_\Ey^2\right)^{1/2} \label{eq:deltadefn}
\end{align}
\end{subequations}
are the longitudinal and transverse components due to fields of
$\mathrm{A}_1$ and $\mathrm{E}$ symmetry, respectively.  From
\eqnref{eq:SymmetrizedFields}, it is clear that a local electric field can
cancel the transverse components of intrinsic strain to restore $C_{3v}$
symmetry to the system, and from \eqnref{eq:meanfdefn} we see that an
electric field $F_z$ applied along the NV-center symmetry axis shifts the
energy of both transitions by the same amount.

We first investigate these effects using device A, shown in
\figref{fig:LateralExpt}(c), consisting of four Ti-Pt-Au gates fabricated on
the diamond surface.  The sample is a 0.5-\milli\metre-thick single-crystal
diamond grown by chemical vapor deposition with $<$5~ppb nitrogen content
(ElementSix), irradiated with \unit{2}{\mega\electronvolt} electrons
(\unit{\numprint{1.2e14}}{\centi\metre\rpsquared}) and then annealed at
\unit{800}{\celsius} to create NV centers. Measurements are performed in a
continuous-flow cryostat operating at $\approx$\unit{20}{\kelvin}. Symmetric
biases $V_X$ and $V_Y$ applied as shown in \figref{fig:LateralExpt}(c)
produce lateral electric fields $F_X$ and $F_Y$ in the $[110]$ and
$[\bar{1}10]$ crystal directions, respectively, while a common dc bias
generates fields in the $[001]$ out-of-plane ($Z$) direction. The sample
$(X,Y,Z)$ and NV-center $(x,y,z)$ coordinate systems are uniquely related for
a given NV-center projection from the $\langle 111\rangle$ family \cite{SOM}.

Figure~\ref{fig:LateralExpt}d contains a series of PLE spectra showing the
optical resonances of the NV center marked in \figref{fig:LateralExpt}(c), as
a function of separate biases $V_X$ and $V_Y$ with $\Vdc=\unit{0}{\volt}$.
The lateral biases are applied as symmetrically pulsed square waves at
\unit{1}{\kilo\hertz}, with PLE photons binned according to polarity. While
the response to static lateral bias is qualitatively similar, this technique
separates slow photoinduced charging effects from the dielectric response as
discussed below.  Two features are evident in the data: first, we observe an
unexpected ``kink'' at zero bias, and second, the resonances cross at
$V_X\approx\unit{7}{\volt}$, demonstrating that we can indeed restore
$C_{3v}$ symmetry to the system.  We explore both of these features below
with additional experiments.

The kink at zero bias reflects an asymmetry in the local electric field
vector as a function of polarity, i.e., $\mathbf{F}(+V)\neq-\mathbf{F}(-V)$.
We argue that this asymmetry results from the photoionization of charge traps
in the diamond host. Even in high-quality single-crystal synthetic diamonds,
deep defects such as vacancy complexes and substitutional nitrogen have
important effects on the material's electronic properties \cite{Isberg2006}.
In particular, substitutional nitrogen atoms form donor levels
$\approx$1.7--\unit{2.2}{\electronvolt} below the conduction band edge
\cite{Farrer1969,*Rosa1999}, and the timescale for charge transport through
these levels is very long (hours) even in nitrogen-rich diamond at room
temperature \cite{Heremans2009}.  These traps are easily ionized by the
\unit{532}{\nano\metre} repump beam ($\approx$\unit{100}{\micro\watt}) which
is 4--5 orders of magnitude stronger than the red laser
($\approx$\unit{1}{\nano\watt}). When voltages are applied, this leads to a
long-lived nonequilibrium charge distribution in the illuminated volume of
the sample, which can either amplify or screen the local electric field.

\begin{figure}[]
\includegraphics{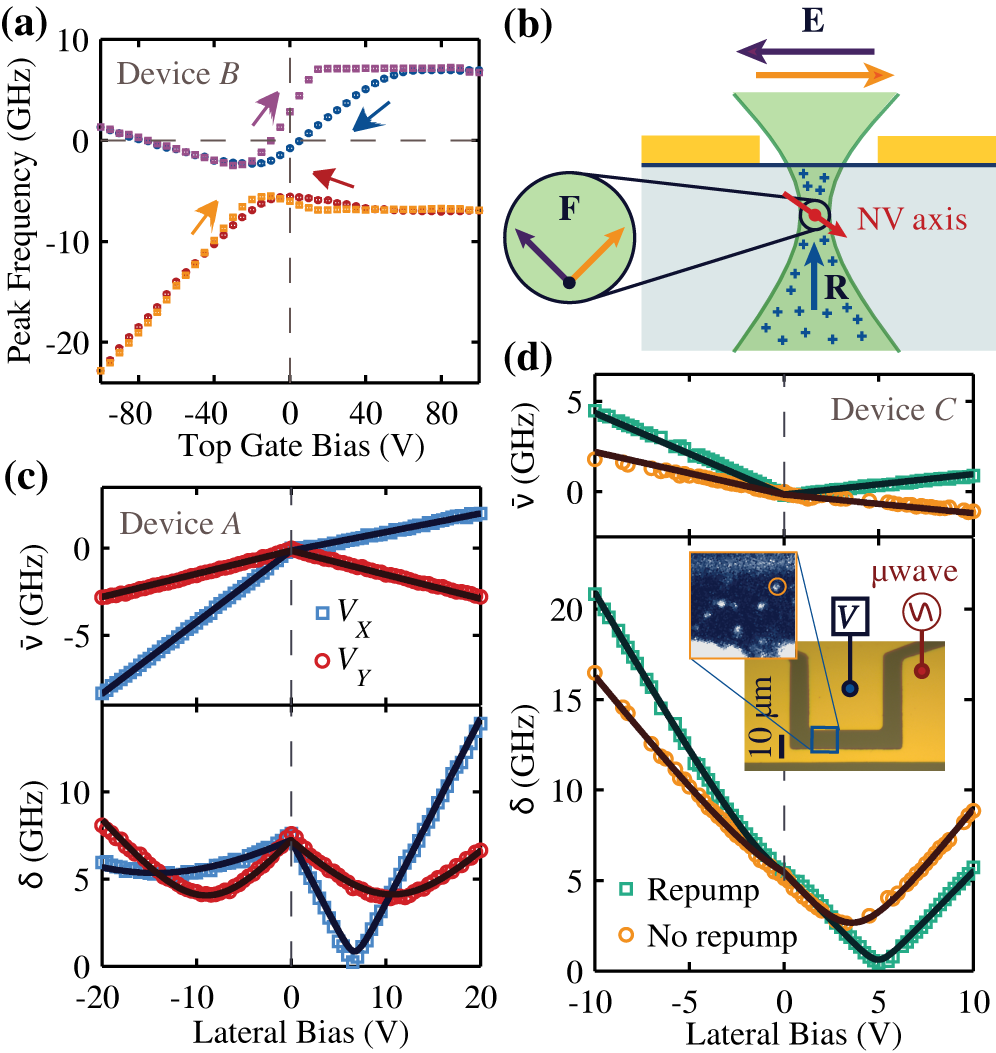}
  \caption[]{\label{fig:Rectification}
(a) Stark-shift hysteresis loop for an NV center \unit{13}{\micro\metre}
below a transparent top gate, as a function of top-gate voltage. Points mark
the transition frequencies from a two-Lorentzian fit to a PLE spectrum at the
corresponding voltage and color-coded arrows indicate the sweep direction.
(b) Schematic of the local electric fields in a lateral geometry.
Photoionized charge traps in the illuminated volume contribute a rectified
field $\mathbf{R}$ predominantly in the $+Z$ direction, which adds to the
dielectric field $\mathbf{E}$ to shift the direction of the local field
$\mathbf{F}$. (c) dc Stark components $\meanf$ and $\delta$ (points)
extracted from fits of the PLE spectra in \figref{fig:LateralExpt}(d), with a
combined fit according to the model described in the text (solid curves). (d)
dc Stark components (points) and fits (solid curves) measured both with
(green squares) and without (orange circles) the \unit{532}{\nano\metre}
repump excitation. Inset: Micrograph and photoluminescence image of device C,
with electrical connections marked.  The 7-\micro\metre-deep NV
center measured in (d) is circled. In all cases, marker sizes slightly exceed
measurement uncertainties.}
\end{figure}

As a simplified one-dimensional demonstration, we present in
\figref{fig:Rectification}(a) the Stark-shift response of an NV center
\unit{13}{\micro\metre} below the surface of a second diamond sample
irradiated and annealed under similar conditions, but patterned with a global
top gate of the transparent conductor indium-tin-oxide (device B).  As the
top-gate bias is stepped in a loop over $\approx$\unit{160}{\minute}, we
observe hysteresis in the response characteristic of $\approx$\unit{1}{\hour}
charge-relaxation timescales. Furthermore, by comparing the magnitudes of the
Stark shifts due to biases applied laterally across an \unit{8}{\micro\metre}
gap [\figref{fig:LateralExpt}(d)] and vertically across the
\unit{0.5}{\milli\metre} sample thickness [\figref{fig:Rectification}(a)], we
find that, when the top-gate bias is negative, the local electric field below
the top gate appears to be amplified by roughly an order of magnitude over
dielectric predictions \cite{SOM}; conversely, the field appears to be
completely screened above a threshold bias, where the response is flat. The
response of an NV center in device A to variations of $\Vdc$ is qualitatively
similar \cite{SOM}, and both are consistent with a picture in which positive
charges in the illuminated volume below the NV center rectify the $Z$
component of the electric field.

We can incorporate these charging effects into a phenomenological model
capturing the essential features of our observations.  As depicted in
\figref{fig:Rectification}(b), the local electric field for an NV center
between two surface gates is composed of a dielectric component roughly
parallel to the sample surface and a rectified component due to photoionized
charge that is mainly out of plane.
We model this field as
\begin{equation}
  \label{eq:LocalField}
  \mathbf{F}= \beta V\dvec + \beta\abs{V}\rectvec,
\end{equation}
where $\beta$ accounts for geometric and dielectric factors that predict a
local electric field in the direction $\dvec$ in response to an applied
voltage $V$, and $\rectvec$ is a dimensionless vector giving the relative
strength and direction of the rectified field, assumed to scale linearly with
$\abs{V}$. Because of the long charging time scale, the rectified field
$\beta\abs{V}\rectvec$ does not change when we switch the bias polarity on
millisecond time scales while the dielectric component $\beta V\dvec$ changes
sign, producing a polarity-asymmetric response. The assumption of a linear
relationship between the rectified field strength and $\abs{V}$ is motivated
by the empirical observation that $\meanf$, proportional to $F_z$, varies
linearly with applied bias in all our measurements. This amounts to an
approximation that the spatial distribution of photoionized charge remains
fixed, while the charge density varies linearly with $\abs{V}$.

Figure \ref{fig:Rectification}(c) shows the mean ($\meanf$) and difference
($\delta$) of the transition frequencies extracted from fits to the PLE
spectra in \figref{fig:LateralExpt}(d). The NV-center symmetry axis
($\left[11\bar{1}\right]$ in this case) is uniquely determined by the sign of
$\meanf$ in response to electric fields in different directions. By
substituting \eqnref{eq:LocalField} into \eqnsref{eq:DCStarkComponents} and
applying the appropriate coordinate transformation, we obtain a model that
quantitatively agrees with our observations \cite{SOM}, as shown by the fits
to the data in \figref{fig:Rectification}(c). Given that this is only a
simplified phenomenological description of a complicated three-dimensional
system, it matches our observations surprisingly well.

Finally, we present a control experiment in which we mitigate effects due to
the \unit{532}{\nano\metre} repump cycle. Occasional repump pulses are still
required to compensate for photoionization of the NV$^-$ charge state due to
sequential two-photon absorption, but with weak ($<$\unit{1}{\nano\watt})
resonant light, the required repump period can be increased to several
seconds \cite{Fu2009}, allowing time to apply bias, record a complete PLE
spectrum, and rezero the bias, all between repump pulses. Since weak
spin-nonconserving optical transitions quickly polarize the NV-center spin
away from resonance in the absence of the repump cycle, we mix the spin
population by applying a microwave magnetic field resonant with the GS spin
transition.  We use another device for this purpose (device C), shown in
\figref{fig:Rectification}(d), which is fabricated on the same diamond as our
four-gate lateral device.  It consists of a short-terminated waveguide to
generate microwave fields and serve as ground, and a gate that when biased
produces lateral electric fields across an \unit{8}{\micro\metre} gap.

Figure \ref{fig:Rectification}(d) shows $\meanf$ and $\delta$ as a function
of gate voltage for the NV center circled in the inset.  Once again the bias
polarity is switched at \unit{1}{\kilo\hertz} and the PLE photons are binned
accordingly. A polarity asymmetry is clearly observed when the repump beam is
present, particularly as a kink in $\meanf$, and it is significantly reduced
when the biases are applied in the absence of the repump cycle. Fits to the
data using our model \cite{SOM} are shown as solid curves, from which we find
that $\abs{\rectvec}$ is reduced from 0.71$\pm$0.02 with the standard repump
cycle to 0.33$\pm$0.03 when the \unit{532}{\nano\metre} beam is omitted.

\begin{figure}[]
\includegraphics{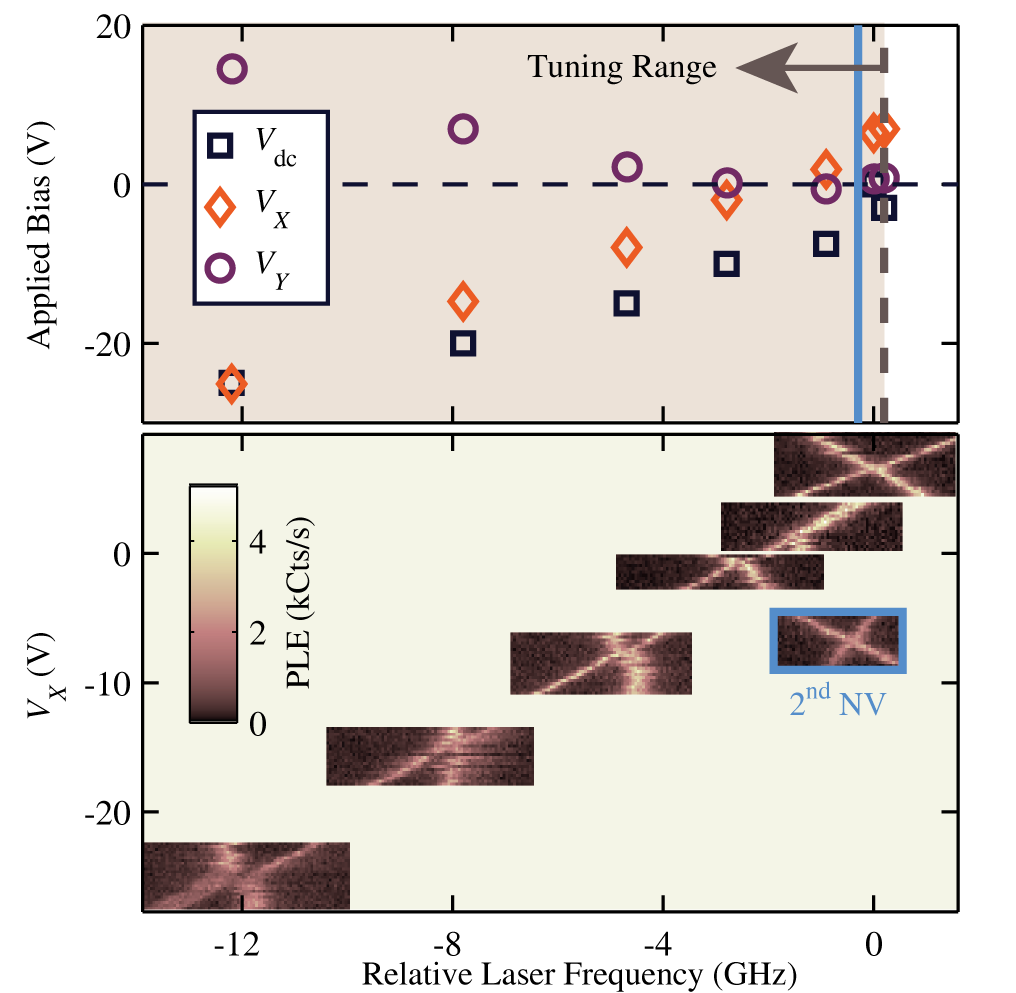}
  \caption[]{\label{fig:Degeneracy}
Tuning diagram for the NV center marked in \figref{fig:LateralExpt}(c).
Degeneracy is achieved at different frequencies by setting the applied biases
$(\Vdc,V_X,V_Y)$ as marked in the upper panel; in the lower panel we show PLE
spectra as a function of $V_X$ around each of these points for fixed $\Vdc$
($V_Y$ is also varied to keep the ratio $V_X/V_Y$ constant). We can shift a
second NV center to degeneracy within the tuning range of the first, as shown
by the PLE spectra outlined in blue (lower panel) and the corresponding blue
line (upper panel) marking the degenerate frequency. }
\end{figure}

Based on this understanding, we can exploit the photoinduced charge to obtain
greatly enhanced tunability in our four-gate geometry (device A). Since the
rectified field points predominantly out of plane and has strength comparable
to the dielectric component, we effectively obtain three-dimensional control
of the local electric field vector.  The application of a negative reference
bias $\Vdc$ as shown in \figref{fig:LateralExpt}(c) increases the rectified
component $F_Z$ independently of $(F_X,F_Y)$.  As a demonstration of the
flexibility of this technique, we present a tuning diagram in
\figref{fig:Degeneracy} in which we use $(V_X,V_Y)$ to scan through the
$C_{3v}$ degeneracy point at different settings of $\Vdc$.
Each crossing occurs at a different frequency, with the corresponding bias
point $(V_X,V_Y,\Vdc)$ marked in the upper panel. Essentially, we are
compensating the transverse components $(S_\Ex,S_\Ey)$ of the intrinsic
fields and tuning the longitudinal component of $\mathbf{F}$ to shift the
frequency. Since the rectified field always points along $+Z$, we can only
tune the frequency in one direction, but the effect is strong enough to
produce a $>$\unit{10}{\giga\hertz} shift in the degenerate frequency with
practical applied voltages.

With this technique we can tune multiple NV centers to have the same
degenerate transition frequency.  The PLE spectra outlined in blue in
\figref{fig:Degeneracy} were obtained from a second NV center in the same
device at $\Vdc=\unit{0}{\volt}$ and display $C_{3v}$ degeneracy at a
frequency within the tuning range of the first.  If these two NV centers were
in separately controlled devices and tuned simultaneously to degeneracy at
the same frequency, they would couple identically to indistinguishable
photons.

In conclusion, we have used electric fields to tune the ZPL transitions of
individual NV centers in micron-scale devices size compatible with photonic
structures. Through their dc Stark shifts, NV centers serve as nanoscale
probes of their electrostatic environment, revealing strong signatures of
charge accumulation due to photoionization of deep donor levels in the
diamond.  We have analyzed these effects with a phenomenological model and
used the additional fields provided by photoionization to obtain
three-dimensional control of the local electric field, in order to tune both
the overall energy and orbital splitting of the excited-state Hamiltonian. In
particular, we have demonstrated how to reach the $C_{3v}$ symmetry point and
then apply longitudinal perturbations to shift the degenerate photon energy.
By coupling multiple NV centers to indistinguishable photons with this
technique, photonic networks could provide a quantum bus to coherently couple
distant NV centers, and entanglement swapping protocols
\cite{Barrett2005,*Childress2005} could enable long-distance quantum key
distribution.

\begin{acknowledgments}
We acknowledge financial support from the AFOSR, ARO, and DARPA, and thank
R.\ Hanson, C.\ G.\ Van de Walle, U.\ K.\ Mishra, K.\ Ohno, and D.\ J.\
Christle for useful discussions.
\end{acknowledgments}

\bibliographystyle{apsrev4-1_CompAuList} 
\bibliography{C:/Users/Lee/Research/References/NVdatabase}

\begin{thebibliography}{27}%
\makeatletter
\providecommand \@ifxundefined [1]{%
 \@ifx{#1\undefined}
}%
\providecommand \@ifnum [1]{%
 \ifnum #1\expandafter \@firstoftwo
 \else \expandafter \@secondoftwo
 \fi
}%
\providecommand \@ifx [1]{%
 \ifx #1\expandafter \@firstoftwo
 \else \expandafter \@secondoftwo
 \fi
}%
\providecommand \natexlab [1]{#1}%
\providecommand \enquote  [1]{``#1''}%
\providecommand \bibnamefont  [1]{#1}%
\providecommand \bibfnamefont [1]{#1}%
\providecommand \citenamefont [1]{#1}%
\providecommand \href@noop [0]{\@secondoftwo}%
\providecommand \href [0]{\begingroup \@sanitize@url \@href}%
\providecommand \@href[1]{\@@startlink{#1}\@@href}%
\providecommand \@@href[1]{\endgroup#1\@@endlink}%
\providecommand \@sanitize@url [0]{\catcode `\\12\catcode `\$12\catcode
  `\&12\catcode `\#12\catcode `\^12\catcode `\_12\catcode `\%12\relax}%
\providecommand \@@startlink[1]{}%
\providecommand \@@endlink[0]{}%
\providecommand \url  [0]{\begingroup\@sanitize@url \@url }%
\providecommand \@url [1]{\endgroup\@href {#1}{\urlprefix }}%
\providecommand \urlprefix  [0]{URL }%
\providecommand \Eprint [0]{\href }%
\providecommand \doibase [0]{http://dx.doi.org/}%
\providecommand \selectlanguage [0]{\@gobble}%
\providecommand \bibinfo  [0]{\@secondoftwo}%
\providecommand \bibfield  [0]{\@secondoftwo}%
\providecommand \translation [1]{[#1]}%
\providecommand \BibitemOpen [0]{}%
\providecommand \bibitemStop [0]{}%
\providecommand \bibitemNoStop [0]{.\EOS\space}%
\providecommand \EOS [0]{\spacefactor3000\relax}%
\providecommand \BibitemShut  [1]{\csname bibitem#1\endcsname}%
\let\auto@bib@innerbib\@empty
\bibitem [{\citenamefont {Balasubramanian}\ \emph {et~al.}(2009)\citenamefont
  {Balasubramanian}, \citenamefont {Neumann}, \citenamefont {Twitchen},
  \citenamefont {Markham} \emph {et~al.}}]{Balasubramanian2009}%
  \BibitemOpen
  \bibfield  {author} {\bibinfo {author} {\bibfnamefont {G.}~\bibnamefont
  {Balasubramanian}} \emph {et~al.},\ }\href
  {http://dx.doi.org/10.1038/nmat2420} {\bibfield  {journal} {\bibinfo
  {journal} {Nat. Mater.}\ }\textbf {\bibinfo {volume} {8}},\ \bibinfo {pages}
  {383} (\bibinfo {year} {2009})}\BibitemShut {NoStop}%
\bibitem [{\citenamefont {Fuchs}\ \emph {et~al.}(2009)\citenamefont {Fuchs},
  \citenamefont {Dobrovitski}, \citenamefont {Toyli}, \citenamefont {Heremans}
  \emph {et~al.}}]{Fuchs2009}%
  \BibitemOpen
  \bibfield  {author} {\bibinfo {author} {\bibfnamefont {G.~D.}\ \bibnamefont
  {Fuchs}} \emph {et~al.},\ }\href {\doibase 10.1126/science.1181193}
  {\bibfield  {journal} {\bibinfo  {journal} {Science}\ }\textbf {\bibinfo
  {volume} {326}},\ \bibinfo {pages} {1520} (\bibinfo {year}
  {2009})}\BibitemShut {NoStop}%
\bibitem [{\citenamefont {Manson}\ \emph {et~al.}(2006)\citenamefont {Manson},
  \citenamefont {Harrison},\ and\ \citenamefont {Sellars}}]{Manson2006}%
  \BibitemOpen
  \bibfield  {author} {\bibinfo {author} {\bibfnamefont {N.~B.}\ \bibnamefont
  {Manson}}, \bibinfo {author} {\bibfnamefont {J.~P.}\ \bibnamefont
  {Harrison}}, \ and\ \bibinfo {author} {\bibfnamefont {M.~J.}\ \bibnamefont
  {Sellars}},\ }\href {\doibase 10.1103/PhysRevB.74.104303} {\bibfield
  {journal} {\bibinfo  {journal} {Phys. Rev. B}\ }\textbf {\bibinfo {volume}
  {74}},\ \bibinfo {eid} {104303} (\bibinfo {year} {2006})}\BibitemShut
  {NoStop}%
\bibitem [{\citenamefont {Fu}\ \emph {et~al.}(2009)\citenamefont {Fu},
  \citenamefont {Santori}, \citenamefont {Barclay}, \citenamefont {Rogers}
  \emph {et~al.}}]{Fu2009}%
  \BibitemOpen
  \bibfield  {author} {\bibinfo {author} {\bibfnamefont {K.-M.~C.}\
  \bibnamefont {Fu}} \emph {et~al.},\ }\href
  {http://link.aps.org/doi/10.1103/PhysRevLett.103.256404} {\bibfield
  {journal} {\bibinfo  {journal} {Phys. Rev. Lett.}\ }\textbf {\bibinfo
  {volume} {103}},\ \bibinfo {pages} {256404} (\bibinfo {year}
  {2009})}\BibitemShut {NoStop}%
\bibitem [{\citenamefont {Buckley}\ \emph {et~al.}(2010)\citenamefont
  {Buckley}, \citenamefont {Fuchs}, \citenamefont {Bassett},\ and\
  \citenamefont {Awschalom}}]{Buckley2010}%
  \BibitemOpen
  \bibfield  {author} {\bibinfo {author} {\bibfnamefont {B.~B.}\ \bibnamefont
  {Buckley}}, \bibinfo {author} {\bibfnamefont {G.~D.}\ \bibnamefont {Fuchs}},
  \bibinfo {author} {\bibfnamefont {L.~C.}\ \bibnamefont {Bassett}}, \ and\
  \bibinfo {author} {\bibfnamefont {D.~D.}\ \bibnamefont {Awschalom}},\ }\href
  {http://www.sciencemag.org/content/330/6008/1212.abstract} {\bibfield
  {journal} {\bibinfo  {journal} {Science}\ }\textbf {\bibinfo {volume}
  {330}},\ \bibinfo {pages} {1212} (\bibinfo {year} {2010})}\BibitemShut
  {NoStop}%
\bibitem [{\citenamefont {Togan}\ \emph {et~al.}(2010)\citenamefont {Togan},
  \citenamefont {Chu}, \citenamefont {Trifonov}, \citenamefont {Jiang} \emph
  {et~al.}}]{Togan2010}%
  \BibitemOpen
  \bibfield  {author} {\bibinfo {author} {\bibfnamefont {E.}~\bibnamefont
  {Togan}} \emph {et~al.},\ }\href {http://dx.doi.org/10.1038/nature09256}
  {\bibfield  {journal} {\bibinfo  {journal} {Nature}\ }\textbf {\bibinfo
  {volume} {466}},\ \bibinfo {pages} {730} (\bibinfo {year}
  {2010})}\BibitemShut {NoStop}%
\bibitem [{\citenamefont {Faraon}\ \emph {et~al.}(2011)\citenamefont {Faraon},
  \citenamefont {Barclay}, \citenamefont {Santori}, \citenamefont {Fu} \emph
  {et~al.}}]{Faraon2011}%
  \BibitemOpen
  \bibfield  {author} {\bibinfo {author} {\bibfnamefont {A.}~\bibnamefont
  {Faraon}} \emph {et~al.},\ }\href {http://dx.doi.org/10.1038/nphoton.2011.52}
  {\bibfield  {journal} {\bibinfo  {journal} {Nat. Photon.}\ }\textbf {\bibinfo
  {volume} {5}},\ \bibinfo {pages} {301} (\bibinfo {year} {2011})}\BibitemShut
  {NoStop}%
\bibitem [{\citenamefont {Santori}\ \emph {et~al.}(2010)\citenamefont
  {Santori}, \citenamefont {Barclay}, \citenamefont {Fu}, \citenamefont
  {Beausoleil} \emph {et~al.}}]{Santori2010}%
  \BibitemOpen
  \bibfield  {author} {\bibinfo {author} {\bibfnamefont {C.}~\bibnamefont
  {Santori}} \emph {et~al.},\ }\href
  {http://stacks.iop.org/0957-4484/21/i=27/a=274008} {\bibfield  {journal}
  {\bibinfo  {journal} {Nanotechnology}\ }\textbf {\bibinfo {volume} {21}},\
  \bibinfo {pages} {274008} (\bibinfo {year} {2010})}\BibitemShut {NoStop}%
\bibitem [{\citenamefont {Barrett}\ and\ \citenamefont
  {Kok}(2005)}]{Barrett2005}%
  \BibitemOpen
  \bibfield  {author} {\bibinfo {author} {\bibfnamefont {S.~D.}\ \bibnamefont
  {Barrett}}\ and\ \bibinfo {author} {\bibfnamefont {P.}~\bibnamefont {Kok}},\
  }\href {\doibase 10.1103/PhysRevA.71.060310} {\bibfield  {journal} {\bibinfo
  {journal} {Phys. Rev. A}\ }\textbf {\bibinfo {volume} {71}},\ \bibinfo
  {pages} {060310} (\bibinfo {year} {2005})}\BibitemShut {NoStop}%
\bibitem [{\citenamefont {Childress}\ \emph {et~al.}(2005)\citenamefont
  {Childress}, \citenamefont {Taylor}, \citenamefont {S{\o}rensen},\ and\
  \citenamefont {Lukin}}]{Childress2005}%
  \BibitemOpen
  \bibfield  {author} {\bibinfo {author} {\bibfnamefont {L.}~\bibnamefont
  {Childress}}, \bibinfo {author} {\bibfnamefont {J.~M.}\ \bibnamefont
  {Taylor}}, \bibinfo {author} {\bibfnamefont {A.~S.}\ \bibnamefont
  {S{\o}rensen}}, \ and\ \bibinfo {author} {\bibfnamefont {M.~D.}\ \bibnamefont
  {Lukin}},\ }\href {http://link.aps.org/doi/10.1103/PhysRevA.72.052330}
  {\bibfield  {journal} {\bibinfo  {journal} {Phys. Rev. A}\ }\textbf {\bibinfo
  {volume} {72}},\ \bibinfo {pages} {052330} (\bibinfo {year}
  {2005})}\BibitemShut {NoStop}%
\bibitem [{\citenamefont {Maze}\ \emph {et~al.}(2008)\citenamefont {Maze},
  \citenamefont {Stanwix}, \citenamefont {Hodges}, \citenamefont {Hong} \emph
  {et~al.}}]{Maze2008}%
  \BibitemOpen
  \bibfield  {author} {\bibinfo {author} {\bibfnamefont {J.~R.}\ \bibnamefont
  {Maze}} \emph {et~al.},\ }\href {http://dx.doi.org/10.1038/nature07279}
  {\bibfield  {journal} {\bibinfo  {journal} {Nature}\ }\textbf {\bibinfo
  {volume} {455}},\ \bibinfo {pages} {644} (\bibinfo {year}
  {2008})}\BibitemShut {NoStop}%
\bibitem [{\citenamefont {Balasubramanian}\ \emph {et~al.}(2008)\citenamefont
  {Balasubramanian}, \citenamefont {Chan}, \citenamefont {Kolesov},
  \citenamefont {Al-Hmoud} \emph {et~al.}}]{Balasubramanian2008}%
  \BibitemOpen
  \bibfield  {author} {\bibinfo {author} {\bibfnamefont {G.}~\bibnamefont
  {Balasubramanian}} \emph {et~al.},\ }\href
  {http://dx.doi.org/10.1038/nature07278} {\bibfield  {journal} {\bibinfo
  {journal} {Nature}\ }\textbf {\bibinfo {volume} {455}},\ \bibinfo {pages}
  {648} (\bibinfo {year} {2008})}\BibitemShut {NoStop}%
\bibitem [{\citenamefont {Maertz}\ \emph {et~al.}(2010)\citenamefont {Maertz},
  \citenamefont {Wijnheijmer}, \citenamefont {Fuchs}, \citenamefont
  {Nowakowski} \emph {et~al.}}]{Maertz2010}%
  \BibitemOpen
  \bibfield  {author} {\bibinfo {author} {\bibfnamefont {B.~J.}\ \bibnamefont
  {Maertz}} \emph {et~al.},\ }\href {\doibase 10.1063/1.3337096} {\bibfield
  {journal} {\bibinfo  {journal} {Appl. Phys. Lett.}\ }\textbf {\bibinfo
  {volume} {96}},\ \bibinfo {eid} {092504} (\bibinfo {year}
  {2010})}\BibitemShut {NoStop}%
\bibitem [{\citenamefont {Dolde}\ \emph {et~al.}(2011)\citenamefont {Dolde},
  \citenamefont {Fedder}, \citenamefont {Doherty}, \citenamefont {Nobauer}
  \emph {et~al.}}]{Dolde2011}%
  \BibitemOpen
  \bibfield  {author} {\bibinfo {author} {\bibfnamefont {F.}~\bibnamefont
  {Dolde}} \emph {et~al.},\ }\href {http://dx.doi.org/10.1038/nphys1969}
  {\bibfield  {journal} {\bibinfo  {journal} {Nat. Phys.}\ }\textbf {\bibinfo
  {volume} {7}},\ \bibinfo {pages} {459} (\bibinfo {year} {2011})}\BibitemShut
  {NoStop}%
\bibitem [{\citenamefont {Batalov}\ \emph {et~al.}(2009)\citenamefont
  {Batalov}, \citenamefont {Jacques}, \citenamefont {Kaiser}, \citenamefont
  {Siyushev} \emph {et~al.}}]{Batalov2009}%
  \BibitemOpen
  \bibfield  {author} {\bibinfo {author} {\bibfnamefont {A.}~\bibnamefont
  {Batalov}} \emph {et~al.},\ }\href
  {http://link.aps.org/doi/10.1103/PhysRevLett.102.195506} {\bibfield
  {journal} {\bibinfo  {journal} {Phys. Rev. Lett.}\ }\textbf {\bibinfo
  {volume} {102}},\ \bibinfo {pages} {195506} (\bibinfo {year}
  {2009})}\BibitemShut {NoStop}%
\bibitem [{\citenamefont {Van~Oort}\ and\ \citenamefont
  {Glasbeek}(1990)}]{VanOort1990}%
  \BibitemOpen
  \bibfield  {author} {\bibinfo {author} {\bibfnamefont {E.}~\bibnamefont
  {Van~Oort}}\ and\ \bibinfo {author} {\bibfnamefont {M.}~\bibnamefont
  {Glasbeek}},\ }\href {\doibase DOI: 10.1016/0009-2614(90)85665-Y} {\bibfield
  {journal} {\bibinfo  {journal} {Chem. Phys. Lett.}\ }\textbf {\bibinfo
  {volume} {168}},\ \bibinfo {pages} {529 } (\bibinfo {year}
  {1990})}\BibitemShut {NoStop}%
\bibitem [{\citenamefont {Tamarat}\ \emph {et~al.}(2006)\citenamefont
  {Tamarat}, \citenamefont {Gaebel}, \citenamefont {Rabeau}, \citenamefont
  {Khan} \emph {et~al.}}]{Tamarat2006a}%
  \BibitemOpen
  \bibfield  {author} {\bibinfo {author} {\bibfnamefont {P.}~\bibnamefont
  {Tamarat}} \emph {et~al.},\ }\href
  {http://link.aps.org/doi/10.1103/PhysRevLett.97.083002} {\bibfield  {journal}
  {\bibinfo  {journal} {Phys. Rev. Lett.}\ }\textbf {\bibinfo {volume} {97}},\
  \bibinfo {pages} {083002} (\bibinfo {year} {2006})}\BibitemShut {NoStop}%
\bibitem [{\citenamefont {Tamarat}\ \emph {et~al.}(2008)\citenamefont
  {Tamarat}, \citenamefont {Manson}, \citenamefont {Harrison}, \citenamefont
  {McMurtrie} \emph {et~al.}}]{Tamarat2008}%
  \BibitemOpen
  \bibfield  {author} {\bibinfo {author} {\bibfnamefont {P.}~\bibnamefont
  {Tamarat}} \emph {et~al.},\ }\href
  {http://stacks.iop.org/1367-2630/10/i=4/a=045004} {\bibfield  {journal}
  {\bibinfo  {journal} {New J. Phys.}\ }\textbf {\bibinfo {volume} {10}},\
  \bibinfo {pages} {045004} (\bibinfo {year} {2008})}\BibitemShut {NoStop}%
\bibitem [{\citenamefont {Doherty}\ \emph {et~al.}(2011)\citenamefont
  {Doherty}, \citenamefont {Manson}, \citenamefont {Delaney},\ and\
  \citenamefont {Hollenberg}}]{Doherty2011}%
  \BibitemOpen
  \bibfield  {author} {\bibinfo {author} {\bibfnamefont {M.~W.}\ \bibnamefont
  {Doherty}}, \bibinfo {author} {\bibfnamefont {N.~B.}\ \bibnamefont {Manson}},
  \bibinfo {author} {\bibfnamefont {P.}~\bibnamefont {Delaney}}, \ and\
  \bibinfo {author} {\bibfnamefont {L.~C.~L.}\ \bibnamefont {Hollenberg}},\
  }\href {http://stacks.iop.org/1367-2630/13/i=2/a=025019} {\bibfield
  {journal} {\bibinfo  {journal} {New J. Phys.}\ }\textbf {\bibinfo {volume}
  {13}},\ \bibinfo {pages} {025019} (\bibinfo {year} {2011})}\BibitemShut
  {NoStop}%
\bibitem [{\citenamefont {Maze}\ \emph {et~al.}(2011)\citenamefont {Maze},
  \citenamefont {Gali}, \citenamefont {Togan}, \citenamefont {Chu} \emph
  {et~al.}}]{Maze2011}%
  \BibitemOpen
  \bibfield  {author} {\bibinfo {author} {\bibfnamefont {J.~R.}\ \bibnamefont
  {Maze}} \emph {et~al.},\ }\href
  {http://stacks.iop.org/1367-2630/13/i=2/a=025025} {\bibfield  {journal}
  {\bibinfo  {journal} {New J. Phys.}\ }\textbf {\bibinfo {volume} {13}},\
  \bibinfo {pages} {025025} (\bibinfo {year} {2011})}\BibitemShut {NoStop}%
\bibitem [{Note1()}]{Note1}%
  \BibitemOpen
  \bibinfo {note} {The NV-center coordinate system is chosen such that
  $\protect \mathaccentV {hat}05E{z}$ points along the N-V symmetry axis and
  $\protect \mathaccentV {hat}05E{x}$ lies in a reflection plane.}\BibitemShut
  {Stop}%
\bibitem [{Note2()}]{Note2}%
  \BibitemOpen
  \bibinfo {note} {Note that while spin-spin coupling leads to mixed ES spin
  eigenstates in some regimes, it does not significantly affect the optical
  pumping mechanism which polarizes the spin into $\left | S_z \right >$ in our
  experiments.}\BibitemShut {Stop}%
\bibitem [{SOM()}]{SOM}%
  \BibitemOpen
  \href@noop {} {}\bibinfo {note} {See supporting online material.}\BibitemShut
  {Stop}%
\bibitem [{\citenamefont {Isberg}\ \emph {et~al.}(2006)\citenamefont {Isberg},
  \citenamefont {Tajani},\ and\ \citenamefont {Twitchen}}]{Isberg2006}%
  \BibitemOpen
  \bibfield  {author} {\bibinfo {author} {\bibfnamefont {J.}~\bibnamefont
  {Isberg}}, \bibinfo {author} {\bibfnamefont {A.}~\bibnamefont {Tajani}}, \
  and\ \bibinfo {author} {\bibfnamefont {D.~J.}\ \bibnamefont {Twitchen}},\
  }\href {http://link.aps.org/doi/10.1103/PhysRevB.73.245207} {\bibfield
  {journal} {\bibinfo  {journal} {Phys. Rev. B}\ }\textbf {\bibinfo {volume}
  {73}},\ \bibinfo {pages} {245207} (\bibinfo {year} {2006})}\BibitemShut
  {NoStop}%
\bibitem [{\citenamefont {Farrer}(1969)}]{Farrer1969}%
  \BibitemOpen
  \bibfield  {author} {\bibinfo {author} {\bibfnamefont {R.~G.}\ \bibnamefont
  {Farrer}},\ }\href {\doibase DOI: 10.1016/0038-1098(69)90593-6} {\bibfield
  {journal} {\bibinfo  {journal} {Solid State Commun.}\ }\textbf {\bibinfo
  {volume} {7}},\ \bibinfo {pages} {685 } (\bibinfo {year} {1969})}\BibitemShut
  {NoStop}%
\bibitem [{\citenamefont {Rosa}\ \emph {et~al.}(1999)\citenamefont {Rosa},
  \citenamefont {Van\v{e}\v{c}ek}, \citenamefont {Nesl\'{a}dek},\ and\
  \citenamefont {Stals}}]{Rosa1999}%
  \BibitemOpen
  \bibfield  {author} {\bibinfo {author} {\bibfnamefont {J.}~\bibnamefont
  {Rosa}}, \bibinfo {author} {\bibfnamefont {M.}~\bibnamefont
  {Van\v{e}\v{c}ek}}, \bibinfo {author} {\bibfnamefont {M.}~\bibnamefont
  {Nesl\'{a}dek}}, \ and\ \bibinfo {author} {\bibfnamefont {L.~M.}\
  \bibnamefont {Stals}},\ }\href
  {http://www.sciencedirect.com/science/article/pii/S0925963598003549}
  {\bibfield  {journal} {\bibinfo  {journal} {Diam. Relat. Mater.}\ }\textbf
  {\bibinfo {volume} {8}},\ \bibinfo {pages} {721} (\bibinfo {year}
  {1999})}\BibitemShut {NoStop}%
\bibitem [{\citenamefont {Heremans}\ \emph {et~al.}(2009)\citenamefont
  {Heremans}, \citenamefont {Fuchs}, \citenamefont {Wang}, \citenamefont
  {Hanson} \emph {et~al.}}]{Heremans2009}%
  \BibitemOpen
  \bibfield  {author} {\bibinfo {author} {\bibfnamefont {F.~J.}\ \bibnamefont
  {Heremans}} \emph {et~al.},\ }\href {\doibase 10.1063/1.3120225} {\bibfield
  {journal} {\bibinfo  {journal} {Appl. Phys. Lett.}\ }\textbf {\bibinfo
  {volume} {94}},\ \bibinfo {eid} {152102} (\bibinfo {year}
  {2009})}\BibitemShut {NoStop}%
\end{thebibliography}%


\begin{thebibliography}{6}%
\makeatletter
\providecommand \@ifxundefined [1]{%
 \@ifx{#1\undefined}
}%
\providecommand \@ifnum [1]{%
 \ifnum #1\expandafter \@firstoftwo
 \else \expandafter \@secondoftwo
 \fi
}%
\providecommand \@ifx [1]{%
 \ifx #1\expandafter \@firstoftwo
 \else \expandafter \@secondoftwo
 \fi
}%
\providecommand \natexlab [1]{#1}%
\providecommand \enquote  [1]{``#1''}%
\providecommand \bibnamefont  [1]{#1}%
\providecommand \bibfnamefont [1]{#1}%
\providecommand \citenamefont [1]{#1}%
\providecommand \href@noop [0]{\@secondoftwo}%
\providecommand \href [0]{\begingroup \@sanitize@url \@href}%
\providecommand \@href[1]{\@@startlink{#1}\@@href}%
\providecommand \@@href[1]{\endgroup#1\@@endlink}%
\providecommand \@sanitize@url [0]{\catcode `\\12\catcode `\$12\catcode
  `\&12\catcode `\#12\catcode `\^12\catcode `\_12\catcode `\%12\relax}%
\providecommand \@@startlink[1]{}%
\providecommand \@@endlink[0]{}%
\providecommand \url  [0]{\begingroup\@sanitize@url \@url }%
\providecommand \@url [1]{\endgroup\@href {#1}{\urlprefix }}%
\providecommand \urlprefix  [0]{URL }%
\providecommand \Eprint [0]{\href }%
\providecommand \doibase [0]{http://dx.doi.org/}%
\providecommand \selectlanguage [0]{\@gobble}%
\providecommand \bibinfo  [0]{\@secondoftwo}%
\providecommand \bibfield  [0]{\@secondoftwo}%
\providecommand \translation [1]{[#1]}%
\providecommand \BibitemOpen [0]{}%
\providecommand \bibitemStop [0]{}%
\providecommand \bibitemNoStop [0]{.\EOS\space}%
\providecommand \EOS [0]{\spacefactor3000\relax}%
\providecommand \BibitemShut  [1]{\csname bibitem#1\endcsname}%
\let\auto@bib@innerbib\@empty
\bibitem [{\citenamefont {Buckley}\ \emph {et~al.}(2010)\citenamefont
  {Buckley}, \citenamefont {Fuchs}, \citenamefont {Bassett},\ and\
  \citenamefont {Awschalom}}]{Buckley2010}%
  \BibitemOpen
  \bibfield  {author} {\bibinfo {author} {\bibfnamefont {B.~B.}\ \bibnamefont
  {Buckley}}, \bibinfo {author} {\bibfnamefont {G.~D.}\ \bibnamefont {Fuchs}},
  \bibinfo {author} {\bibfnamefont {L.~C.}\ \bibnamefont {Bassett}}, \ and\
  \bibinfo {author} {\bibfnamefont {D.~D.}\ \bibnamefont {Awschalom}},\ }\href
  {http://www.sciencemag.org/content/330/6008/1212.abstract} {\bibfield
  {journal} {\bibinfo  {journal} {Science}\ }\textbf {\bibinfo {volume}
  {330}},\ \bibinfo {pages} {1212} (\bibinfo {year} {2010})}\BibitemShut
  {NoStop}%
\bibitem [{\citenamefont {M\"{u}ller}\ \emph {et~al.}(2011)\citenamefont
  {M\"{u}ller}, \citenamefont {Aharonovich}, \citenamefont {Lombez},
  \citenamefont {Alaverdyan}, \citenamefont {Vamivakas}, \citenamefont
  {Castelletto}, \citenamefont {Jelezko}, \citenamefont {Wrachtrup},
  \citenamefont {Prawer},\ and\ \citenamefont {Atat\"{u}re}}]{Muller2011}%
  \BibitemOpen
  \bibfield  {author} {\bibinfo {author} {\bibfnamefont {T.}~\bibnamefont
  {M\"{u}ller}}, \bibinfo {author} {\bibfnamefont {I.}~\bibnamefont
  {Aharonovich}}, \bibinfo {author} {\bibfnamefont {L.}~\bibnamefont {Lombez}},
  \bibinfo {author} {\bibfnamefont {Y.}~\bibnamefont {Alaverdyan}}, \bibinfo
  {author} {\bibfnamefont {A.~N.}\ \bibnamefont {Vamivakas}}, \bibinfo {author}
  {\bibfnamefont {S.}~\bibnamefont {Castelletto}}, \bibinfo {author}
  {\bibfnamefont {F.}~\bibnamefont {Jelezko}}, \bibinfo {author} {\bibfnamefont
  {J.}~\bibnamefont {Wrachtrup}}, \bibinfo {author} {\bibfnamefont
  {S.}~\bibnamefont {Prawer}}, \ and\ \bibinfo {author} {\bibfnamefont
  {M.}~\bibnamefont {Atat\"{u}re}},\ }\href
  {http://stacks.iop.org/1367-2630/13/i=7/a=075001} {\bibfield  {journal}
  {\bibinfo  {journal} {New J. Phys.}\ }\textbf {\bibinfo {volume} {13}},\
  \bibinfo {pages} {075001} (\bibinfo {year} {2011})}\BibitemShut {NoStop}%
\bibitem [{\citenamefont {Nebel}\ \emph {et~al.}(1998)\citenamefont {Nebel},
  \citenamefont {Stutzmann}, \citenamefont {Lacher}, \citenamefont {Koidl},\
  and\ \citenamefont {Zachai}}]{Nebel1998}%
  \BibitemOpen
  \bibfield  {author} {\bibinfo {author} {\bibfnamefont {C.~E.}\ \bibnamefont
  {Nebel}}, \bibinfo {author} {\bibfnamefont {M.}~\bibnamefont {Stutzmann}},
  \bibinfo {author} {\bibfnamefont {F.}~\bibnamefont {Lacher}}, \bibinfo
  {author} {\bibfnamefont {P.}~\bibnamefont {Koidl}}, \ and\ \bibinfo {author}
  {\bibfnamefont {R.}~\bibnamefont {Zachai}},\ }\href
  {http://www.sciencedirect.com/science/article/pii/S0925963597002033}
  {\bibfield  {journal} {\bibinfo  {journal} {Diam. Relat. Mater.}\ }\textbf
  {\bibinfo {volume} {7}},\ \bibinfo {pages} {556} (\bibinfo {year}
  {1998})}\BibitemShut {NoStop}%
\bibitem [{\citenamefont {Heremans}\ \emph {et~al.}(2009)\citenamefont
  {Heremans}, \citenamefont {Fuchs}, \citenamefont {Wang}, \citenamefont
  {Hanson},\ and\ \citenamefont {Awschalom}}]{Heremans2009}%
  \BibitemOpen
  \bibfield  {author} {\bibinfo {author} {\bibfnamefont {F.~J.}\ \bibnamefont
  {Heremans}}, \bibinfo {author} {\bibfnamefont {G.~D.}\ \bibnamefont {Fuchs}},
  \bibinfo {author} {\bibfnamefont {C.~F.}\ \bibnamefont {Wang}}, \bibinfo
  {author} {\bibfnamefont {R.}~\bibnamefont {Hanson}}, \ and\ \bibinfo {author}
  {\bibfnamefont {D.~D.}\ \bibnamefont {Awschalom}},\ }\href {\doibase
  10.1063/1.3120225} {\bibfield  {journal} {\bibinfo  {journal} {Appl. Phys.
  Lett.}\ }\textbf {\bibinfo {volume} {94}},\ \bibinfo {eid} {152102} (\bibinfo
  {year} {2009})}\BibitemShut {NoStop}%
\bibitem [{\citenamefont {Twitchen}\ \emph {et~al.}(1999)\citenamefont
  {Twitchen}, \citenamefont {Hunt}, \citenamefont {Smart}, \citenamefont
  {Newton},\ and\ \citenamefont {Baker}}]{Twitchen1999}%
  \BibitemOpen
  \bibfield  {author} {\bibinfo {author} {\bibfnamefont {D.~J.}\ \bibnamefont
  {Twitchen}}, \bibinfo {author} {\bibfnamefont {D.~C.}\ \bibnamefont {Hunt}},
  \bibinfo {author} {\bibfnamefont {V.}~\bibnamefont {Smart}}, \bibinfo
  {author} {\bibfnamefont {M.~E.}\ \bibnamefont {Newton}}, \ and\ \bibinfo
  {author} {\bibfnamefont {J.~M.}\ \bibnamefont {Baker}},\ }\href {\doibase
  DOI: 10.1016/S0925-9635(99)00038-2} {\bibfield  {journal} {\bibinfo
  {journal} {Diam. Relat. Mater.}\ }\textbf {\bibinfo {volume} {8}},\ \bibinfo
  {pages} {1572 } (\bibinfo {year} {1999})}\BibitemShut {NoStop}%
\bibitem [{\citenamefont {Kittel}(1996)}]{Kittel1996}%
  \BibitemOpen
  \bibfield  {author} {\bibinfo {author} {\bibfnamefont {C.}~\bibnamefont
  {Kittel}},\ }\href@noop {} {\emph {\bibinfo {title} {Introduction to Solid
  State Physics}}},\ \bibinfo {edition} {7th}\ ed.\ (\bibinfo  {publisher}
  {John Wiley \& Sons, Inc.},\ \bibinfo {year} {1996})\ p.\ \bibinfo {pages}
  {574}\BibitemShut {NoStop}%
\end{thebibliography}%

\end{document}


\title{Supplemental Material for \\
``Electrical Tuning of Single Nitrogen-Vacancy Center\\
Optical Transitions Enhanced by Photoinduced Fields''}

\author{L.\ C.\ Bassett}
\author{F.\ J.\ Heremans}
\author{C.\ G.\ Yale}
\author{B.\ B.\ Buckley}
\author{D.\ D.\ Awschalom}
\affiliation{Center for Spintronics and Quantum Computation,\\ University of
California, Santa Barbara, CA 93106, USA}


\date{\today}
\maketitle

\section{Methods and supplementary data}

\subsection{Optical methods}

Measurements were performed using two home-built confocal microscopes
designed to excite single NV centers both resonantly at
\unit{637.2}{\nano\metre} and in the blue-shifted phonon sideband at
\unit{532}{\nano\metre}, while collecting photoluminescence photons from the
red-shifted photoluminescence sideband between
$\approx$650--\unit{830}{\nano\metre}.  Details and a schematic are available
elsewhere \cite{Buckley2010}. The \unit{532}{\nano\metre} laser serves both
as the excitation for standard photoluminescence spatial imaging and as a
`repump' cycle to maintain the NV$^-$ charge state which is photoionized by
resonant excitation alone.

Resonant excitation is provided by tunable diode lasers from the New Focus
Velocity$^\mathrm{TM}$ series, which are tuned by means of a piezoelectric
actuator that makes fine adjustments to the laser cavity.  An important
technical aspect of these experiments is the calibration of the laser
frequency response to variations in the voltage applied to the piezo, since
this is typically nonlinear and hysteretic.  In one of our microscope setups
we have pre-calibrated the laser frequency response to particular piezo
modulation sequences using a Mach-Zehnder interferometer.  In the other
setup, unpredictable mode hops of the laser make such calibrations
unreliable, so we continuously monitor the laser frequency
interferometrically: a Fabry-Perot cavity with a free spectral range of
\unit{1.5}{\giga\hertz} provides fine frequency resolution, while a
Mach-Zehnder interferometer with free spectral range
$\approx$\unit{7}{\giga\hertz} is used to detect mode hops.  In both cases,
our calibrations provide relative frequency determinations with
$\approx$\unit{100}{\mega\hertz} uncertainty across the
$\approx$\unit{90}{\giga\hertz} piezo modulation range. The resonant light is
prepared with circular polarization, so that the (linear) optical selection
rules for $\ket{S_z}$ optical transitions do not cause the resonances to
disappear in certain regimes.

In both setups, we also have the ability to add confocal excitation at
\unit{405}{\nano\metre} (\unit{3.1}{\electronvolt}) to the optical path.
Unsurprisingly, this higher-energy excitation significantly alters the
charging dynamics of the system.  Generally it causes the NV center optical
transitions to become less stable and so it was not used during any of the
measurements presented in the main text, but we did use it as a control to
`reset' the equilibrium charge distribution with no applied bias between
experiments. A short burst of \unit{405}{\nano\metre} light applied in this
configuration was generally enough to restore the initial NV center
transition frequencies ($\approx$\unit{10}{\second} at optical power
$\approx$\unit{10}{\micro\watt}), while full relaxation in the presence of
the \unit{532}{\nano\metre} light alone required several hours.

\subsection{Charging behavior in device \emph{A}}

\begin{figure}[]
\includegraphics{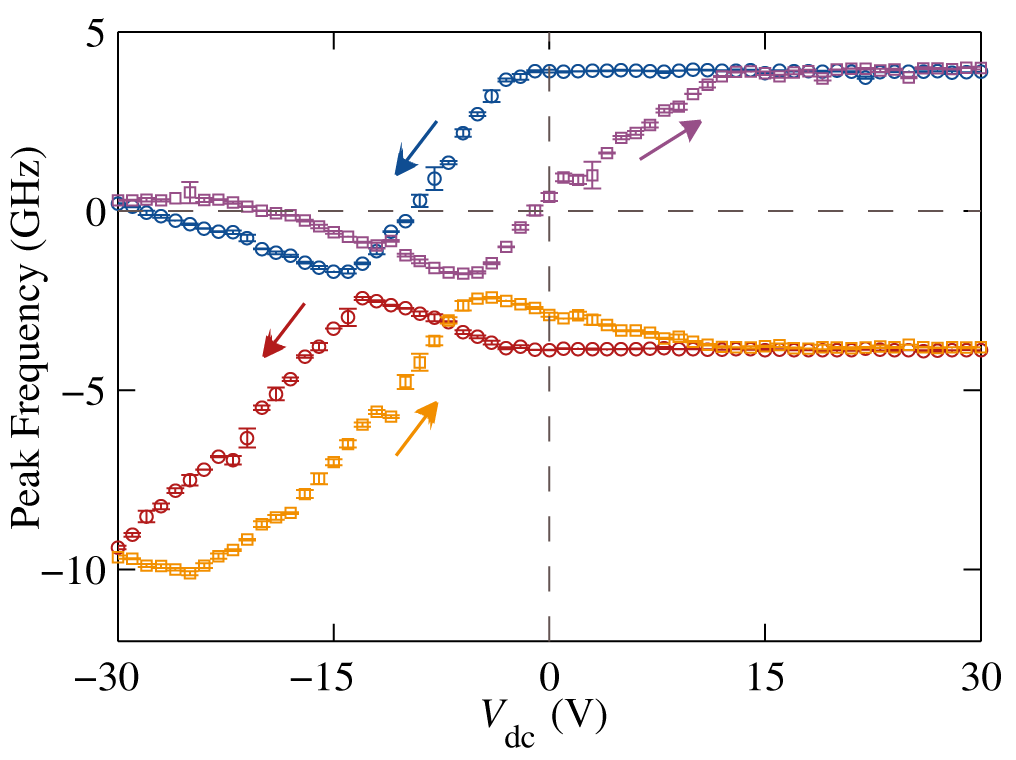}
  \caption[]{\label{fig:LateralHysteresis}
Stark shift hysteresis loop for the NV center marked in Fig.~1c of the main
text as a function of $\Vdc$ (with $V_X,V_Y=0$).  Points mark the optical
resonance frequencies extracted from two-Lorentzian fits to PLE spectra at
the corresponding voltage, and color-coded arrows mark the sweep direction.
Errorbars represent the frequency uncertainties from the fits. }
\end{figure}

The interpretation of the hysteresis measurements presented in the main text
(Fig.~2a) is simplified by the one-dimensional geometry of the top-gate
device, but we measured a qualitatively similar response in our four-gate
lateral device (device \emph{A}, see Fig.~1c in the main text) to variations
of the reference voltage $\Vdc$ common to all four gates.  Figure
\ref{fig:LateralHysteresis} shows a hysteresis loop as a function of $\Vdc$
for the NV center marked in Fig.~1c, which is $\approx$\unit{7}{\micro\metre}
below the diamond surface.
Although the response is similar to the top-gate measurements we do notice a
few differences: First, the hysteresis `direction' is reversed, in the sense
that the Stark response lags $\Vdc$ in \figref{fig:LateralHysteresis}, while
it leads the top-gate bias in Fig.~2a. Second, the `threshold bias' is closer
to zero in the lateral device than in our top-gate measurements. These
differences are probably due to a variety of factors, including (a) geometric
differences due to the gate geometry and the depths of the respective NV
centers, (b) variations in the diamond samples, (c) details of the
diamond-metal interfaces for different metals (Ti/Pt/Au on the lateral device
and indium-tin-oxide as the top gate), and perhaps most importantly (d) the
electrostatic boundary conditions at the back of the device -- the top-gate
device \emph{B} is indium-bonded to ground on the backside, while device
\emph{A} is glued to the ground plane with an insulating adhesive.

We note that polarity-dependent Stark shifts have been recently observed for
Chromium centers in diamond \cite{Muller2011}, and suggest that a
one-dimensional experiment with a transparent top gate could determine
whether these devices display similar charging behavior.

\subsection{Photoconductivity}

\begin{figure}[]
\includegraphics{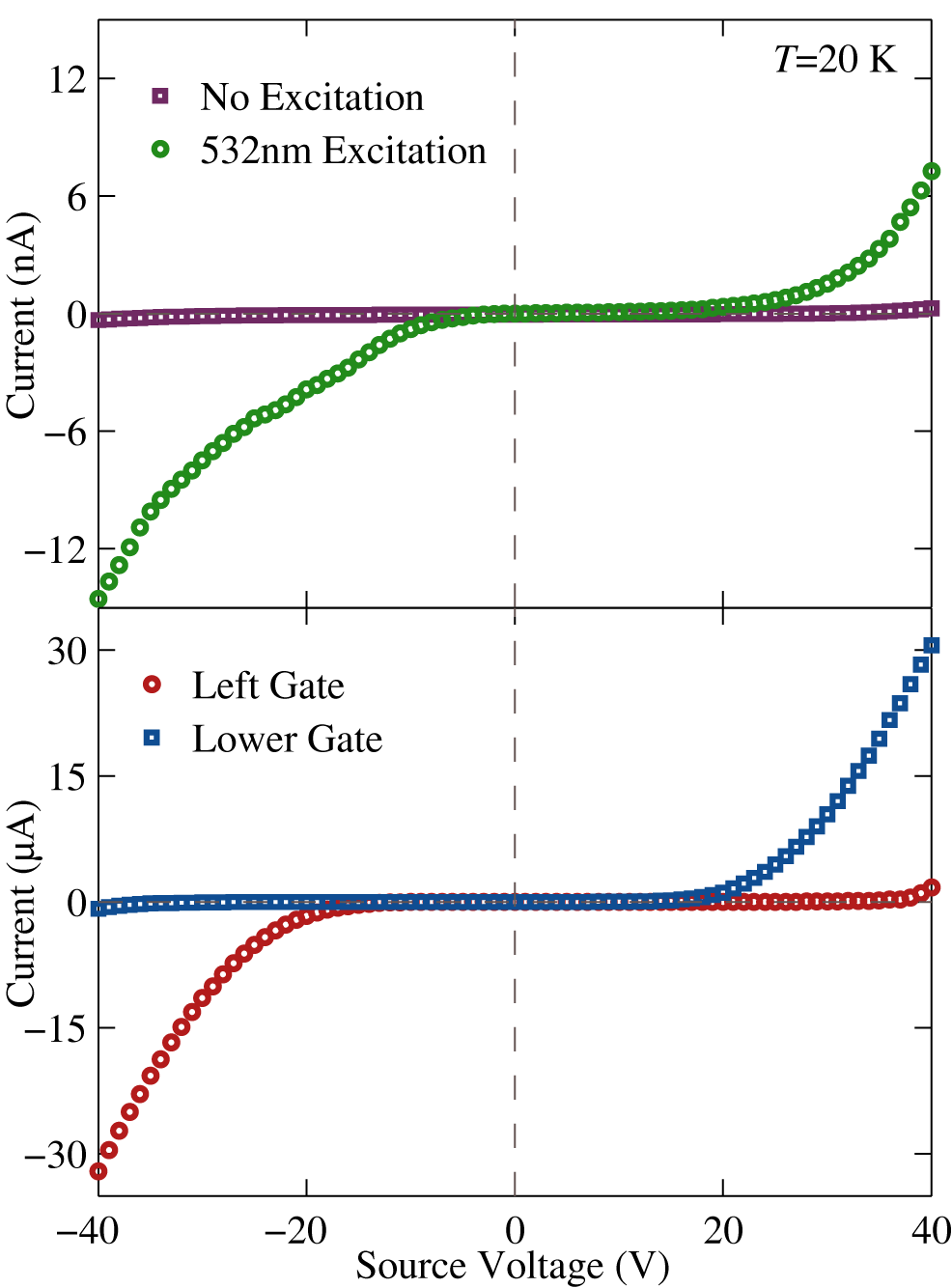}
  \caption[]{\label{fig:Photoconductivity}
Top panel: $I$-$V$ curves showing conductivity between gate the right-hand
gate of device \emph{A} (see Fig.~1c in the main text) and the remaining three
surface gates, both with (green circles) and without (purple squares)
\unit{532}{\nano\metre} excitation.  Lower panel: Similar photoconductivity
measurements (excitation applied) with the bias connected to the left-hand
gate (red circles) and lower gate (blue squares), showing a larger leakage
current between these two gates.}
\end{figure}

When voltages are applied between surface gates in the presence of optical
excitation, we expect the photoexcitation of defect levels to lead to a
photocurrent near the surface.  \figref{fig:Photoconductivity} shows the
photoconductivity response of device \emph{A} when the excitation beam is
focused on the NV center marked in Fig.~1c of the main text,
\unit{6}{\micro\metre} below the diamond surface.
The top panel shows an example of a typical dc photoconductivity response for
this and similar devices we have measured, with almost no measurable current
flowing in the dark and typical photocurrents of a few nanoamperes when the
excitation beam is applied. On this particular device, a larger current
`leak' existed between the left and lower gates, as shown in the lower panel
of \figref{fig:Photoconductivity}. When
$V_\mathrm{Lower}-V_\mathrm{Left}\gtrsim \unit{20}{\volt}$, a current of
several microamperes flows between them.

It is important to note, however, that the presence of a photocurrent between
gates near the diamond surface does not preclude the accumulation of a space
charge deeper in the sample, especially since photoexcited charge transport
through diamond is an inefficient process, and the timescales for charge
storage are very long \cite{Nebel1998,*Heremans2009}.  As illustrated
schematically in Fig.~2b of the main text, the out-of-plane component of the
photoinduced field is primarily due to positive charge from photoionized
donors \emph{below} the NV center, which results from the effective voltage
applied \emph{across the sample} rather than between two lateral gates. Thus
even with a substantial leak between two gates, both the dielectric and
photoinduced components of the dc Stark shift are always present.
Practically, however, photocurrents may lead to local heating of the sample,
and we attribute the leak between the left and lower gates to the broadening
of the PLE resonances observed in Fig.~3 of the main text when larger lateral
voltages are applied.

\section{Analysis of dc Stark shifts}

The first step in the analysis of dc Stark shifts in our devices is a
determination of the NV center's orientation in the diamond crystal.  This
may be any of the eight projections from the $\langle 111\rangle$ family;
note that antiparallel (i.e., N-V compared to V-N) orientations are
distinguishable due to the permanent dipole in the $z$ direction.  Using the
coordinate systems described in the text, we note that N-V axis projects
either in the $\pm X$ or $\pm Y$ sample directions, and also has a component
along $\pm Z$.  We can therefore label a given orientation by
$(p_X,p_Y,p_Z)$, where $p_{X,Y} = 0,\pm 1$ give the sign of the N-V
projection in the $(X,Y)$ axes (note $p_X^2+p_Y^2=1$) and $p_Z = \pm 1$ gives
the sign of the $Z$-projection.  For a given NV center, we can determine
whether $p_X$ or $p_Y$ equals zero from the dependence of the
photoluminescence on the linear polarization of the \unit{532}{\nano\metre}
repump beam (the photoluminescence is a maximized when the light is polarized
perpendicular to the N-V axis). It is then usually straightforward to
determine the exact projection from the slope of the longitudinal dc Stark
component $\meanf$ in response to applied voltages, keeping in mind that the
rectified field points mainly out of plane.  For example, we know from
polarization measurements that the NV center of Fig.~1 projects in the $\pm
X$ direction, and from the sign of $\meanf$ in response to $V_X$ (Fig.~2c) we
see that $p_X=+1$ and $p_Z=-1$, corresponding to the crystal direction
$[11\bar{1}]$ as depicted in the schematic of Fig.~2b.

Given the NV-center symmetry axis, the transformation between sample and
NV-center coordinate systems is given by
\begin{equation}
  \mathbf{F}_\mathrm{NV} = \mathrm{M}\mathbf{F}_\mathrm{Sample},
\end{equation}
where $\mathrm{M}$ is the rotation matrix
\begin{widetext}
\begin{equation}
  \mathrm{M} = \frac{1}{\sqrt{3}}\left(
  \begin{array}{ccc}
p_X & p_Y & -\sqrt{2}p_Z \\
-\sqrt{3}p_Y p_Z & \sqrt{3}p_X p_Z & 0 \\
\sqrt{2}p_X & \sqrt{2}p_Y & p_Z
\end{array}\right).
\end{equation}
Applying this transformation for the NV center in our lateral device, we
obtain expressions for the dc Stark components $\meanf$ and $\delta$ in terms
of the local field in sample coordinates:
\begin{subequations}  \label{eq:StarkComponentsSampleCoords}
\begin{align}
  h\meanf & = \hbar\omega_0 + \frac{\dmupar}{\sqrt{3}}\left(\sqrt{2}F_X - F_Z\right), \\
  h\delta & =
    \sqrt{2}\left\{ \Bigl[S_\Ex - \frac{\mu_\perp}{\sqrt{3}}\left(F_X+\sqrt{2}F_Z\right)\Bigr]^2
              + \left[S_\Ey - \mu_\perp F_Y\right]^2\right\}^{1/2}.
\end{align}
\end{subequations}
\end{widetext}
By expressing the local field according to our model as $\mathbf{F} = \beta
V\dvec + \beta\abs{V}\rectvec$, we obtain a dependence on applied bias that
we can quantitatively compare to our measurements.

To reduce the number of free parameters, we fix the components of the
dielectric unit vector $\dvec$ based on electrostatic simulations of our
device using the \textsc{COMSOL Multiphysics}$^\circledR$ software.  For a
single set of measurements (e.g., $\meanf$ and $\delta$ as a function of bias
along a single direction), this leaves eight free parameters:
$\left\{\omega_0,S_\Ex,S_\Ey,\beta\dmupar,\mu_\perp/\dmupar,\xi_X,\xi_Y,\xi_Z\right\}$.
This equals the combined number of free parameters expected for the form of
both $\meanf$ and $\delta$ (i.e., two lines and two hyperbolae constrained to
meet at $V=0$) and so results in a well-conditioned minimization problem.
When we have multiple measurements of the same NV center, e.g., for the
repump/no repump measurements in Fig.~2d, we can constrain the parameters
even further, and require that
$\left\{\omega_0,S_\Ex,S_\Ey,\beta\dmupar,\mu_\perp/\dmupar\right\}$ are
common for each NV center, while only the rectification vector $\rectvec$
varies independently for each set. In the case of the lateral measurements of
Fig.~2c, we also allow the dielectric coefficient $\beta$ to vary between the
$V_X$ and $V_Y$ measurements, and the resulting best-fit parameters for a
combined fit as described above are listed in Table~\ref{tab:LateralFits}. As
expected, the rectified field points mainly in the $+Z$ direction.

\begin{table*}
\caption{\label{tab:LateralFits}Best-fit parameter values from fits of our
model to the data in Fig.~2c in the main text, for Stark shifts as a function
of applied bias ($V_X,V_Y)$ in orthogonal directions. Uncertainties are at
95\% confidence from the fit alone, and do not include systematic errors due
to the somewhat strong covariance between many of these parameters and the
uncertainty in the fixed direction $\dvec$ of the dielectric field.  Based on
electrostatic modeling, we use $\dvec=(0.998,0.04,0.07)$ and
$(-0.002,0.9998,0.02)$ for the $V_X$ and $V_Y$ dependence, respectively.}
\begin{ruledtabular}
\begin{tabular}{ccccc}
 \multicolumn{2}{c}{\textrm{Common}} & \multicolumn{1}{c}{\textrm{Direction-dependent}} & \multicolumn{2}{c}{\textrm{Bias direction}}\\
 \multicolumn{2}{c}{\textrm{parameters}} & \multicolumn{1}{c}{\textrm{parameters}} & \multicolumn{1}{c}{$V_X$} & \multicolumn{1}{c}{$V_Y$} \\ \hline
 $\omega_0/2\pi$ & $0.66\pm\unit{0.03}{\giga\hertz}$ & $\beta\dmupar/h$ & $0.334\pm\unit{0.002}{\giga\hertz\per\volt}$ & $0.175\pm\unit{0.003}{\giga\hertz\per\volt}$ \\
 $S_\Ex/h$ & $5.06\pm\unit{0.03}{\giga\hertz}$ & $\xi_X$ & $0.15\pm0.01$ & $0.19\pm0.02$ \\
 $S_\Ey/h$ & $0.78\pm\unit{0.06}{\giga\hertz}$ & $\xi_Y$ & $-0.43\pm0.02$ & $-0.30\pm0.02$ \\
 $\mu_\perp/\dmupar$ & $1.41\pm0.02$ & $\xi_Z$ & $1.01\pm0.01$ & $1.60\pm0.02$ \\
\end{tabular}
\end{ruledtabular}
\end{table*}

\subsection{Electrostatics of the photoinduced fields}

A full description of the photoinduced fields in our devices is beyond the
scope of this work, especially since our phenomenological model seems to
capture the behavior important for controlling NV-center Stark shifts in
these geometries. Such a description would presumably need to account for the
gate geometries of our devices, the variation of optical intensity through
the sample in a confocal geometry, transport dynamics of photoexcited
charges, and potentially spatial inhomogeneities in the distribution of deep
donors. Furthermore, although we expect that substitutional nitrogen atoms
dominate the electrostatics, other defect centers may also play an important
role, such as vacancy complexes unintentionally created during the
irradiation and annealing process.  Based on the electron irradiation dose
used, we expect a vacancy concentration (before annealing) of
$\approx$\unit{\numprint{6e13}}{\centi\metre\rpcubed} \cite{Twitchen1999},
which is roughly an order of magnitude less than the nitrogen concentration
in these samples, and these acceptor defects probably have an effect on the
transport of photoionized charge through the sample. Still, we can gather
some quantitative information about the photoinduced charge distributions
from the NV-center Stark response, which we can compare to a simplified
`band-bending' model treating the device as a Schottky diode.

The top-gate geometry of device \emph{B} is the simplest to analyze, and as
mentioned in the text we can compare the magnitudes of the dc Stark shifts
observed in that device (Fig.~2a) with those of device \emph{A} (Fig.~2c) to
estimate the local electric field strength.  Assuming that in the top-gate
device the local electric field points only in the $+Z$ direction, we relate
the slope of $\meanf$ to the local electric field $F_Z = \beta_Z V$ through
the relation
\begin{equation}
  \d{\meanf}{V} = \frac{\beta_Z\dmupar}{\sqrt{3}h},
\end{equation}
where the factor $1/\sqrt{3}$ appears due to the projection of the NV-center
symmetry axis in the sample.  Below the threshold bias in Fig.~2a,
$d\meanf/dV\approx\unit{0.06}{\giga\hertz\per\volt}$. In comparison, from the
fits to the lateral data in Fig.~2c described in the previous section, we
find that the dielectric response of device \emph{A} is
$\beta_\mathrm{Lateral}\dmupar/h\approx \unit{0.2}{\giga\hertz\per\volt}$.
Therefore the ratio of the dielectric field in device \emph{A} to the
rectified field in device \emph{B} is $\beta_\mathrm{Lateral}/\beta_Z\approx
2$, while from dielectric response alone this ratio should be $\approx$100,
corresponding to an enhancement of the vertical electric field below the top
gate by a factor of $\approx$50.

If we picture the top-gate device as a Schottky diode, this comparison
suggests that when negative (reverse) bias is applied, the potential drops
across a distance of only $\approx$\unit{10}{\micro\metre} rather than the
full \unit{0.5}{\milli\metre} thickness of the sample.  Assuming uniform
ionization of deep donors with density $N_d$, the equivalent Schottky
`depletion width' obtained from the solution to Poisson's equation is given
by \cite{Kittel1996}
\begin{equation}
  w_d = \sqrt{\frac{2\epsilon\epsilon_0\abs{\varphi}}{eN_d}},
\end{equation}
in terms of the Schottky barrier height $\varphi$ and diamond's dielectric
constant $\epsilon\approx5.7$.  For $w_d\approx\unit{10}{\micro\metre}$ and
$\varphi\approx\unit{10}{\volt}$, this estimate yields an ionized charge
density of $N_d\approx\unit{\numprint{6e13}}{\centi\metre\rpcubed}$, which is
in agreement with a $\approx$10\% ionization of the anticipated
substitutional nitrogen concentration of a few parts per billion
($\approx$\unit{\numprint{1e15}}{\centi\metre\rpcubed}).  This rough
calculation shows that the photoinduced fields we observe are consistent with
the expected concentration of deep defects, although it does not account for
nonlinearities in the distribution of photoionized charge.  For example, we
observe a qualitatively similar enhancement of the $Z$-direction field when
focussed on NV centers up to $\approx$\unit{50}{\micro\metre} below the top
gate, suggesting that the charge distribution is not confined to a layer near
the surface gate but rather follows the region of higher optical intensity.

\begin{table*}
\caption{\label{tab:RepumpFits}Best-fit parameter values from fits of our
model to the data in Fig.~2d in the main text, investigating the role of the
\unit{532}{\nano\metre} repump laser on the photoinduced field.  As in
Table~\ref{tab:LateralFits}, uncertainties are quoted at 95\% confidence from
the fit, but do not account for covariance between the fit parameters and the
direction of the dielectric field, fixed to $\dvec = (0.97,0.04,-0.24)$ based
on electrostatic modeling.}
\begin{ruledtabular}
\begin{tabular}{ccccc}
 \multicolumn{2}{c}{\textrm{Common}} & \multicolumn{1}{c}{\textrm{Repump-dependent}} & &  \\
 \multicolumn{2}{c}{\textrm{parameters}} & \multicolumn{1}{c}{\textrm{parameters}} & \multicolumn{1}{c}{\textrm{With repump}} & \multicolumn{1}{c}{\textrm{Without repump}} \\ \hline
 $\omega_0/2\pi$ & $-0.17\pm\unit{0.02}{\giga\hertz}$ & $\xi_X$ & $-0.239\pm0.009$ & $0.06\pm0.01$ \\
 $S_\Ex/h$ & $-2.34\pm\unit{0.09}{\giga\hertz}$ & $\xi_Y$ & $0.12\pm0.01$ & $0.13\pm0.02$ \\
 $S_\Ey/h$ & $3.08\pm\unit{0.05}{\giga\hertz}$ & $\xi_Z$ & $0.66\pm0.02$ & $0.30\pm0.03$ \\
 $\mu_\perp/\dmupar$ & $0.91\pm0.01$ & & & \\
 $\beta\dmupar/h$ & $0.99\pm\unit{0.01}{\giga\hertz\per\volt}$ & & & \\
\end{tabular}
\end{ruledtabular}
\end{table*}

\subsection{The role of the repump cycle on photoinduced fields}

\begin{figure}
\includegraphics[width=3.2in]{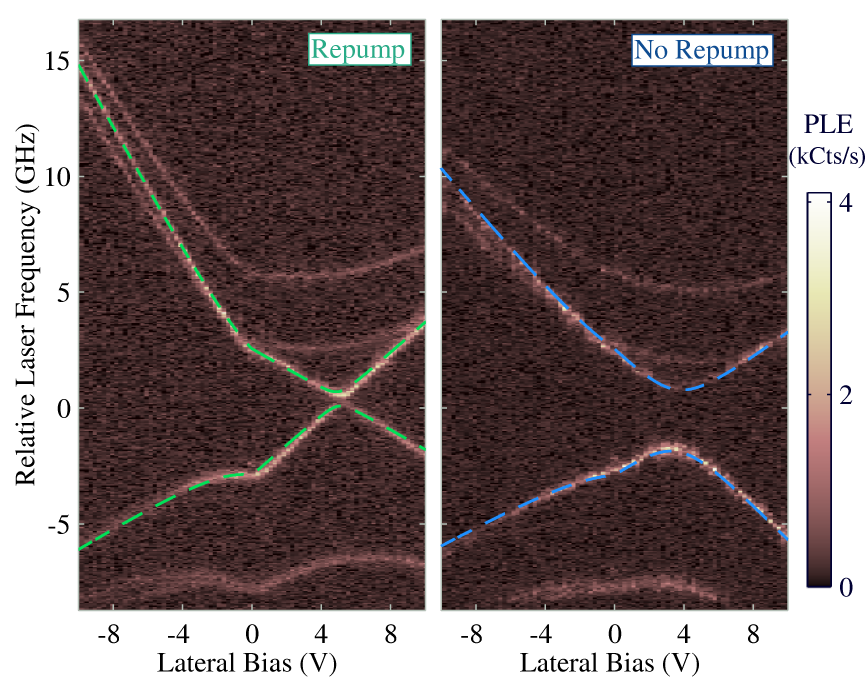}
  \caption[]{\label{fig:GreenOnOff}
PLE spectra as a function of gate voltage, for the NV center circled in
Fig.~2d, both with the standard \unit{532}{\nano\metre} repump cycle (left)
and with only intermittent repump pulses applied at zero bias (right).  The
$\ket{S_z}$ transition frequencies reconstructed from the fits shown in
Fig.~2d are overlaid (dashed curves).  Other resonances visible in these
spectra correspond to spin-conserving optical transitions for the
$\left\{\ket{S_x},\ket{S_y}\right\}$ spin states, as described in the text. }
\end{figure}

As described in the main text, we performed a control experiment to
investigate the influence of the \unit{532}{\nano\metre} repump cycle on the
rectified component of the local electric field.  Using the device shown in
Fig.~2d, we apply a microwave current through a short-terminated waveguide
resonant with the ground-state
$\ket{S_z}\Leftrightarrow\left\{\ket{S_x},\ket{S_y}\right\}$ transition at
\unit{2.878}{\giga\hertz}.  This maintains a mixed spin state during our
measurements since, in the absence of the repump cycle, weak
spin-nonconserving transitions would otherwise polarize the spin away from
resonance with the red laser. In the PLE spectra (\figref{fig:GreenOnOff}),
we therefore observe the spin-conserving optical transitions for all three
triplet spin states. We isolate the $\ket{S_z}$ resonance frequencies from
fits to these PLE spectra to compute the dc Stark components $\meanf$ and
$\delta$ plotted in Fig.~2d of the main text.  These are marked by dashed
curves in \figref{fig:GreenOnOff}, which are the resonance frequencies
reconstructed from the best-fit results of our model plotted in Fig.~2d. The
analysis leading to the fits is similar to that described above (in this case
the NV-center projection is $[1\bar{1}1]$), and the best-fit parameters are
listed in Table~\ref{tab:RepumpFits}.

\bibliography{C:/Users/Lee/Research/References/NVdatabase}